\documentclass[doublecol,figures]{epl2} 

\usepackage{graphicx,color}

\usepackage{amsmath}

\def\rmd{{\rm d}}
\def\rme{{\rm e}}
\def\e{\epsilon}
\def\t0{\tau_0}

\newcommand{\beq}{\begin{equation}}
\newcommand{\eeq}{\end{equation}}
\newcommand{\bea}{\begin{eqnarray}}
\newcommand{\eea}{\end{eqnarray}}
\newcommand{\HH}{{\mathcal H}}
\newcommand\sn{{\rm sn}}
\newcommand\dn{{\rm dn}}

\newcommand\cn{{\rm cn}}

\newcommand\bU{{\mathcal B}}
\newcommand{\reff}[1]{(\ref{#1})}

\title{Quantum quenches as classical critical films}
\author{Andrea Gambassi\inst{1} \and Pasquale Calabrese\inst{2}}
\shortauthor{A. Gambassi and P. Calabrese}

\institute{                    
  \inst{1}SISSA -- International School for Advanced Studies and INFN, via Bonomea 265, 34136  Trieste, Italy\\
  \inst{2}Dipartimento di Fisica dell'Universit\`a di Pisa and INFN, Pisa, Italy
}
\pacs{64.70.Tg}{Quantum phase transitions}
\pacs{64.60.Ht}{Dynamic critical phenomena}
\pacs{05.70.Jk}{Critical point phenomena}

\abstract{
We study the unitary time evolution of the order parameter of a quantum system 
after a sudden quench in the parameter which drives the transition. 
By mapping the dynamics onto  the imaginary time path integral 
in a film geometry we derive the full mean-field {\it non-equilibrium phase diagram} for a one-component order parameter. 
The recently discovered non-equilibrium transition is identified with the shifted critical point 
in films and therefore it is generally expected to occur in more than one spatial dimension.
We also find that anharmonic oscillations of the order parameter are a general feature of the mean-field dynamics 
after the quench.
}

\begin{document}

\maketitle

\section{Introduction}
The experimental realization \cite{uc} of
many-body cold atomic systems evolving in the absence of dissipation triggered a significant 
theoretical activity aimed at understanding the fate of a quantum interacting
system initially prepared in a state which is not a Hamiltonian eigenstate. % 
A global {\it quantum quench} is the prototypical instance: 
the system is prepared at time $t=0$ in the ground state
of a Hamiltonian $H_0$ and then it is let evolve with a Hamiltonian $H$ which differs from $H_0$ {\it everywhere} in space (e.g., due to a sudden quench of an external field). 
The absence of a solid and general theoretical framework 
for studying the ensuing non-equilibrium evolution motivated the
investigation of simple one-dimensional models 
with exactly solvable dynamics (see, e.g., Refs.~\cite{ir-00,cc-06,c-06,cdeo-07,rsms-08},
but this list is far from being exhaustive).  Several aspects of their evolution have been explored both analytically 
and numerically \cite{num}, as reviewed in Ref.~\cite{rev}. 
Only few studies of models in spatial dimension $d>1$ are presently available~\cite{cc-06,dgp-10,sc-10}.
\par

One of the central issues concerning the time evolution after a quench is whether a stationary state exists at long times. It is commonly believed \cite{ir-00,cc-06,cdeo-07,num,gg,scc-09} that, in case it does, it is characterized by correlation functions which resemble those of either a thermal state or a so-called generalized Gibbs ensemble. 
However, a stationary asymptotic state does not always exist since 
the order parameter of the system might display persistent anharmonic oscillations, as it occurs 
in the case of the mean-field (MF) approximation for superfluid bosons \cite{aa-02,sb-10,s-11}, for the             
Hubbard model \cite{hub} and for some exactly solvable pairing models  
\cite{bl-06,dyac-07,fcc-09}. 

A general approach to the study of quench problems, proposed in Ref.~\cite{cc-06}, consists in mapping the real-time evolution of a $d$-dimensional quantum system onto a boundary problem in $d+1$ dimensions in imaginary time. 
With this framework the quantum evolution governed by critical Hamiltonians \cite{cc-06,dis} 
or by massive integrable field theories in $d=1$ \cite{fm-10}, has been successfully investigated.

In this paper we consider such a mapping for a quench within a generic 
effective Landau-Ginzburg (LG) Hamiltonian, which is expected to capture the universal properties of a rather large class of quantum critical points, encompassing some of the cases mentioned above. 
For a one-component order parameter, i.e., within the Ising universality class, 
we present a remarkably simple complete MF solution for the order parameter evolution which, as we shall see, accounts comprehensively for all the qualitative features observed in real-time 
MF analyses\footnote{%
In the case of systems with $n$-component vector order parameter and $O(n)$ symmetry --- such as those  discussed in Refs.~\cite{aa-02,sb-10,s-11,hub,bl-06} ---  the analysis reported here applies to the only component of the order parameter which does not vanish when the $O(n)$ symmetry is broken by the boundaries.%
}.
%%%%%%%%%%%%%%%%%%%%%%%%%%%%%%%%
The main result of our investigation is that the novel "dynamical transition"
(which has no equilibrium counterpart) reported in Refs.~\cite{sb-10,hub} is the manifestation, in real time,  of a rather well-known effect in confined critical systems, i.e., the critical point shift in films \cite{NF}.
Remarkably, the universality of this phenomenon allows us to conclude that the dynamical transition is not a MF 
artifact but
it occurs even in the presence of fluctuations, an important issue which 
was left open by all the analytical approaches considered so far in the literature. 
In addition, we find that anharmonic oscillations are general features of the MF evolution and we  characterize quantitatively their features (form, amplitude, and period) in terms of the quench parameters. 
However, the existence of the transition beyond MF does not guarantee that these oscillations do actually withstand the effects of quantum fluctuations.
Experiments and additional theoretical studies should be able to provide a conclusive answer to this subtle and difficult question.
%
%
%%%%%%%%%%%%%%%%%%%%%%%%%
\section{The mapping to the film geometry}
Following Ref.~\cite{cc-06}, consider a $d$-dimensional
system prepared in a state $|\Psi_0\rangle$ which is not an eigenstate of the 
Hamiltonian $H$ controlling the time evolution. The expectation value of a local operator ${\cal O}(x)$ at time $t$ is given by the path-integral  
\beq
\langle {\cal O}(t,x)\rangle=
Z^{-1} \langle \Psi_0 | \rme^{i H t-\e H} {\cal O}(x) \rme^{-i H t-\e H}|
\Psi_0 \rangle\,,
\label{Oexp}
\eeq
where $Z=\langle\Psi_0|\rme^{-2\e H}|\Psi_0\rangle$ and 
the damping factors $\rme^{-\e H}$ have been introduced in order 
to make the path-integral representation of 
$\langle {\cal O}(t,x)\rangle$ absolutely convergent. Equation~\reff{Oexp} can be represented by
an analytically continued path integral in imaginary time $\tau$ in which the field $\Psi$
takes the boundary values $\Psi_0$ for $\tau=\tau_{1,2}=\pm\e-it$.
The operator ${\cal O}$ is inserted at $\tau=0$. 
The width of the film is $\tau_2-\tau_1 = 2\e$. In the course of the calculation
$\tau_{1,2}$ must be considered as real numbers and only at the
very end they are continued to their effective complex values. 
In this way, the real-time non-equilibrium evolution of a $d$-dimensional system is
mapped onto the thermodynamics of a $d+1$-dimensional field theory in a film
geometry in which the initial state $|\Psi_0\rangle$ plays
the role of a boundary condition at both boundaries.
This approach can be used to study the time-evolution of a generic operator ${\cal O}$. However, in what follows we will concentrate on the order parameter $\phi(t,x)$.

When the system is close to a (quantum) phase transition, we can take advantage of the
renormalization-group (RG) theory of boundary critical phenomena \cite{dd} in order to describe the system in terms of the LG Hamiltonian for the order parameter $\phi(\tau,x)$ in a film  \cite{dd}:
\bea
\HH[\phi] &= &
\int \rmd^d x \int_{\tau_1}^{\tau_2} \rmd\tau 
\left[
\frac{1}{2}(\vec{\partial} \phi)^2+\frac{r}{2}
\phi^2+\frac{g}{4!}\phi^4\right] \nonumber  \\ &&
+\int \rmd^d x \, \left[ \bU(\phi(\tau_1,x)) +
  \bU(\phi(\tau_2,x))\right]\,, \label{LGH} 
\eea
[$\vec \partial \equiv (\vec{\nabla},\partial_\tau)$] where  $\bU(\phi)\equiv  \frac{1}{2} c \phi^2 - h_s \phi$
represents the boundary density,
$r$ controls the distance from the critical point, 
and $g>0$ ensures the stability for $r<0$. 
$c$ is related to the surface enhancement, i.e., to 
the difference between the interaction strengths within the boundaries
and within the bulk, whereas $h_s$ is a symmetry-breaking
field acting at both confining surfaces. 
%
%
%
%%%%%%%%%%%%%%%%%%%%%%%%%%%%%%%%%%%%%
%%%%%%%%%%%%%%%%%%%%%%%%%%%%%%%%%%%%%
\begin{figure}[t]
\centerline{\includegraphics[width=0.46\textwidth]{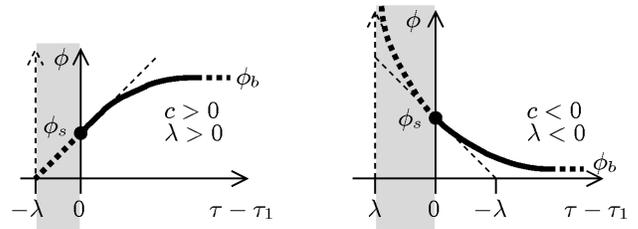}}
\caption{%
Order parameter profiles close to the boundary,
for $h_s=0$ and $c>0$ (left),  $c<0$ (right). 
The RG invariant boundary conditions,
i.e., $\phi_s^*=0$ and $\phi_s^* = \pm\infty$, are effectively
imposed at $\tau = \tau_1 - |\lambda|$. 
For $h_s$ large enough the profile  
looks like the  right one.%
}
\label{fig:surf}
\end{figure}
%%%%%%%%%%%%%%%%%%%%%%%%%%%%%%%%%%%%%%
%%%%%%%%%%%%%%%%%%%%%%%%%%%%%%%%%%%%%%
%
The values of $c$ and $h_s$ determine the actual boundary conditions for the field $\phi$. 
This is easily understood in the semi-infinite geometry $\tau_2\rightarrow\infty$, taking into account the MF boundary condition $- \partial_\tau \phi_s +c \phi_s - h_s = 0$ implied by Eq.~\reff{LGH} for $\phi_s\equiv\phi(\tau_1,x)$, whereas $\phi_b\equiv \phi(\tau\rightarrow\infty,x)$ attains its bulk value $\phi_b = \sqrt{-6r/g}$ for $r<0$, and $\phi_b=0$ for 
$r\ge0$.
For $h_s=0$ and  $\phi_s\neq 0$, $|\phi|$  increases upon leaving the
boundary if $c>0$, as shown by the left panel of Fig.~\ref{fig:surf}. 
The  order parameter profile extrapolated outside the film vanishes  for $\tau = \tau_1- \lambda$ % < \tau_1$ 
(see Fig.~\ref{fig:surf}), where $\lambda$ is the
so-called {\it extrapolation length} $\lambda \simeq 1/c$ 
for $c$ large enough~\cite{dd}. 
If $c<0$, instead, $|\phi|$ decreases upon leaving the surface, as shown by the right panel of
Fig.~\ref{fig:surf}, and the (negative) extrapolation length $\lambda$ determines the 
position $\tau = \tau_1 + \lambda$ at which the MF order parameter profile {\it diverges}.
In order to have $\phi_s\neq 0$ with $h_s=0$, one needs either $r>0$ and $-c>r^{1/2}$ or $r<0$, which correspond to $\phi_b =0$ or $\phi_b\neq 0$, respectively.
A large enough $h_s\neq 0$ causes an enhancement of $|\phi|$ close to the surface, so that $\phi_s$ does not vanish independently of the value of $\phi_b$ and the profile looks like the one depicted on the right panel of Fig.~\ref{fig:surf}.

Accounting for fluctuations requires a RG analysis \cite{dd} which yields the following conclusions 
for translational invariant boundary condition: for $h_s=0$, the possible fixed-point values of $c$ are 
$c^*=\infty,0,-\infty$, with $c^*=\pm\infty$ stable and $c^*=0$ unstable. 
In particular $c^*=+\infty$ leads to Dirichlet (D) boundary condition for $\phi$ at the boundaries of the film, i.e.,
$\phi_s$ flows under  RG towards $\phi_s^*=0$. 
The fixed-point with $c^*=-\infty$ describes both the
cases $(-c> r^{\Phi},h_s=0)$~\cite{dd} and $(c <\infty,h_s\neq 0)$
and leads to a divergence of the order parameter profile upon
approaching the surface: $\phi(\tau\rightarrow 0,x) \simeq
\tau^{-\beta/\nu}$ (with $\beta$ and $\nu$ the standard critical exponents), i.e., $\phi_s$ flows to $\phi_s^* = \pm\infty$, referred to as $\pm$ boundary conditions.
The kind of asymptotic boundary conditions can be also expressed in terms of  
$\phi_s$ and $\phi_b$.
If $\phi_s=0$, the corresponding boundary condition is Dirichlet. 
On the other hand, if $\phi_s\neq 0$, the order parameter profile looks like on the
left or right of Fig.~\ref{fig:surf}, depending on whether $\phi_s<\phi_b$ or
$\phi_s>\phi_b$,  which correspond respectively to D
and $+$ boundary condition.

%%%%%%% 
Going back to the two boundaries problem, the most important feature 
of the phase diagram is the {\it critical point shift} \cite{NF}.
This is easily understood in the case of classical critical points for which $r=T-T_c$, being $T_c$ the critical temperature in the \emph{bulk}. For DD boundary conditions, the suppression of order at the surfaces prevents the film from
ordering and therefore $\phi=0$ everywhere even slightly below $T_c$. 
However, upon further decreasing $T$, the increasingly strong 
tendency towards ordering overcomes the effects of
the boundaries and the whole film orders. This occurs at the shifted critical point 
$T=T_c(L)<T_c$, which depends on the thickness $L$ of the film, and is such that 
$T_c(L) - T_c \sim L^{-1/\nu}$. 
For $++$ boundary conditions, such a point is additionally shifted off-coexistence and it can be reached by
applying a non-zero bulk magnetic field which overcomes the order at the surface.

%%%%%%%%%%%%%%%%%%%%%%%%%%%%%%%%%%%
\par
{\it  Mean-field order-parameter profiles --} Let us consider the LG Hamiltonian (\ref{LGH}) in a
film geometry of thickness $L \equiv \tau_2-\tau_1$ and introduce the rescaled MF order-parameter profile 
$\phi(\tau) = \frac{1}{L} \sqrt{\frac{3!}{g}} \psi((\tau-\tau_1)/L)$, which satisfies
the equation $-\psi'' + \bar r \psi + \psi^3 = 0$,  where $\bar r \equiv rL^2$. For DD boundary conditions,
$\psi(0)=\psi(1)=0$, whereas for $++$ boundary conditions, $\psi(x\to 0)
\simeq x^{-1}$ and $\psi(x\to 1^-) \simeq (1-x)^{-1}$. The differential
equation can be solved analytically. For $++$ boundary conditions and 
$\bar r > \bar r_c=-\pi^2$ (high-temperature phase H) one finds~\cite{krech}
\beq
\psi_{++}(x) = 2\sqrt{2} K  \; 
\displaystyle{\frac{\dn(2 K x;k)}{\sn(2 K x;k)}}\,,
\eeq
where $K \equiv K(k)$ is the complete elliptic integral of the first kind
and the elliptic modulus $k = k(\bar r)$ solves $\bar r = (2 K)^2 (2 k^2 - 1)$. 
$\dn$ and $\sn$ are the standard Jacobi elliptic functions.
For $\bar r \le \bar r_c$ (low-temperature
phase L), instead, one finds \cite{krech}
\beq
\psi_{++}(x) = 2\sqrt{2} K  \; [\sn(2 K x; k)]^{-1}\,,
\eeq
where $k$ satisfies $\bar r = - (2 K)^2 (k^2 + 1)$.

\begin{figure}[t]
\centerline{\includegraphics[width=0.48\textwidth]{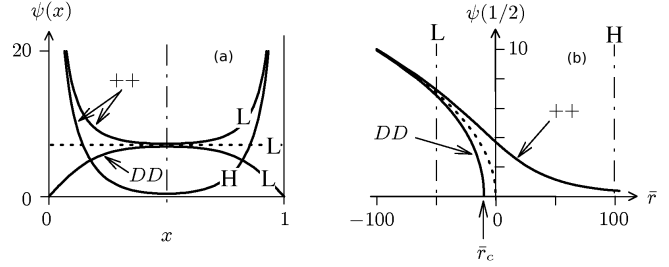}}
\caption{%
(a) Order parameter profiles $\psi(x)$ 
across a film with DD  and $++$ boundary conditions. 
The profiles denoted by L and H correspond, respectively, to  
the low- and high-temperature phases.
(b) Value of the order parameter in the middle of the film as functions of $\bar r$. 
L and H indicate the values of $\bar r$ corresponding to $\psi(x)$  in panel (a).
In both panels, the dotted lines refer to the bulk order parameter.}
\label{fig:profiles}
\end{figure}

For DD boundary conditions, $\psi$ vanishes for $\bar r \ge \bar r_c$.  For $\bar r < \bar r_c$, instead,
the film orders and the order parameter is~\cite{gd}
\beq
\psi_{DD}(x) = 2\sqrt{2} K \; k \,\sn(2 K x; k)\,,
\eeq
where now  $k$ solves  $\bar r =  - (2 K)^2 (k^2 + 1)$.
In Fig.~\ref{fig:profiles}(a) we plot $\psi$ 
for the cases of $++$ and DD boundary conditions. 
%
%%%%%%%%%%%%%%%%%%%%%%%%%%%%%%%%%%%
%%%%%%%%%%%%%%%%%%%%%%%%%%%%%%%%%%% 
%
%
%%%%%%%%%%%%%%%%%%%%%%%%%%%%%%%%%%%%%%%%%
\section{The analytic continuation}
\begin{figure}[t!]
\centerline{\includegraphics[width=0.225\textwidth]{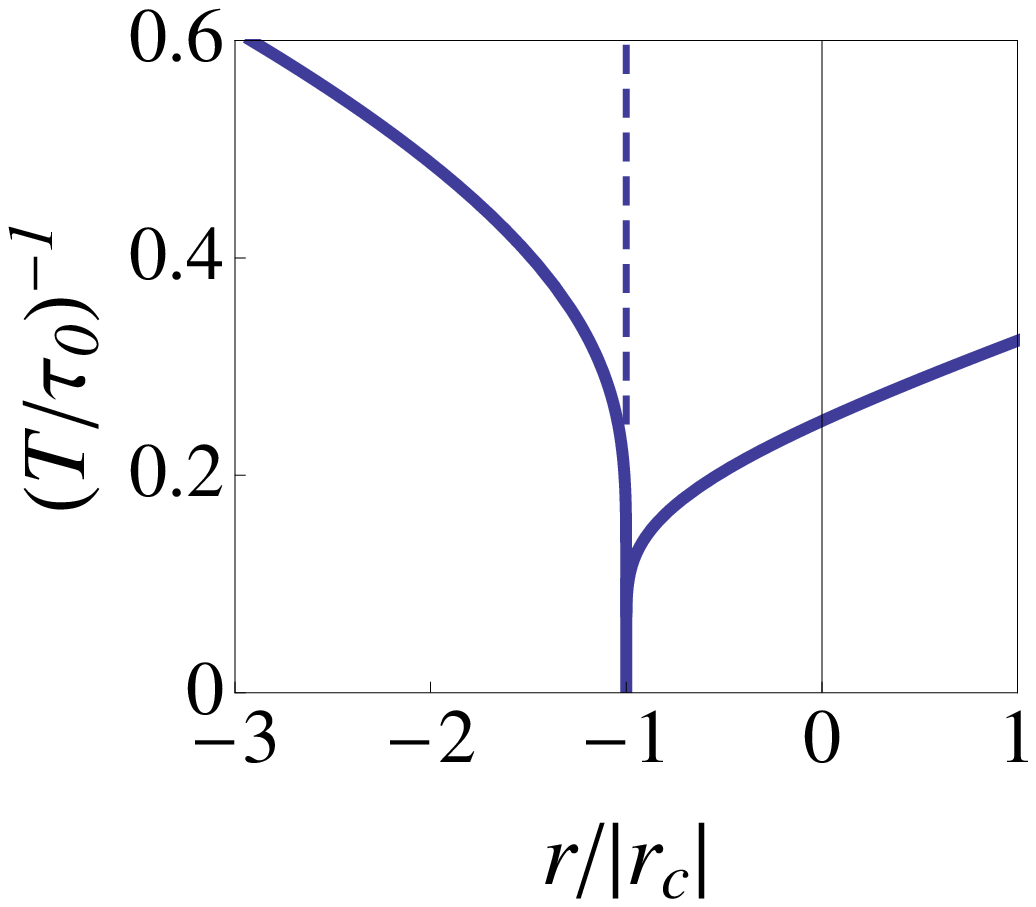}
\includegraphics[width=0.216\textwidth]{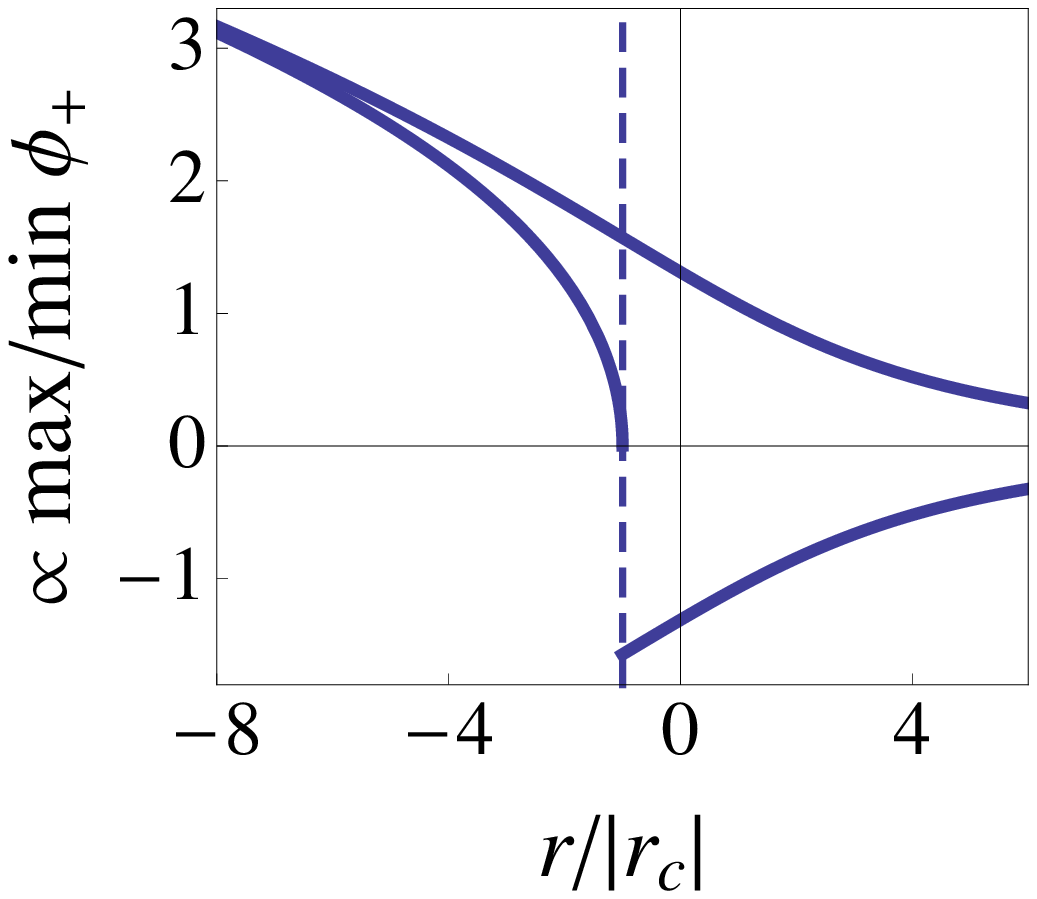}}
\caption{%
Left: Inverse period of the oscillations of $\phi_{+}(t)$ in
unit of $\tau_0^{-1}$ as a function of $r/|r_c|$. 
For $r\rightarrow r_c$ the period diverges logarithmically.
Right: Maximum and minimum value of $\phi_{+}(t)$ (up to the multiplicative constant $a$) as a function of $r/|r_c|$.%
}
\label{fig:ppTA}
\end{figure}

\begin{figure}[t]
\centerline{\includegraphics[width=0.46\textwidth]{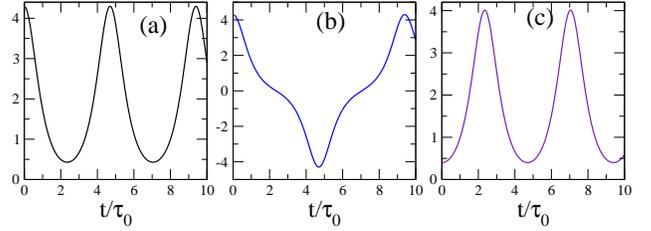}}
\caption{%
Order parameter $\phi$  (in arbitrary units) as a function of $t/\tau_0$:
(a) $\phi_{+}(t)$ for $r<r_c$
 is always positive.
(b) $\phi_{+}(t)$ for $r>r_c$ 
oscillates symmetrically around $0$.
(c) $\phi_{D}(t)$ for $r<r_c$. 
Note that $\phi_{D}(t) = 0$ for $r>r_c$.%
}
\label{fig:prof}
\end{figure}

For the purpose of extracting the asymptotic behavior of $\phi(t)$ for $t\gg\tau_0$ we
replace $|\Psi_0\rangle$ by the appropriate RG-invariant boundary condition $|\Psi_0^*\rangle$ to which it flows. 
The difference is taken into account, to leading order, by assuming that the
RG-invariant boundary conditions are not imposed at $\tau=\tau_{1,2}$
but at $\tau=\tau_{1,2}\mp\tau_0$, where
$\tau_0$ is  the absolute value of the  extrapolation length. 
In the quantum non-equilibrium problem, $\tau_0$ is of the order of the 
correlation length in the initial state \cite{cc-06}. 
The effect of introducing $\tau_0$ is simply to replace $\e$ by
$\e+\tau_0$ and the limit $\e\to0^+$ can now be taken, resulting in a film of
effective width $L=2\tau_0$. 
In order to determine the time evolution of 
the order parameter $\phi$ %at time $t$ 
one has 
simply to continue analytically the expressions provided above for $\psi$ 
to the complex position $\tau = \tau_0 + it$ within the film. Such a continuation is straightforwardly done by using 
standard formulas for the Jacobi functions \cite{elliptic}. 
For a quench with $++$ boundary conditions to a point below $r_c = - \pi^2/(4\tau_0^2)$, one finds
\beq
 \mbox{[L]}\quad
\left\{
\begin{array}{l}
\phi_{+}(t) =  a\; K \,\dn(K t/\tau_0;k')\,, \\
r/|r_c|  = - (2 K/\pi)^2 (k^2 + 1) < -1\,,
\end{array}
\right. 
\label{ppLT}
\eeq
where $k' \equiv \sqrt{1-k^2}$ and
$a\equiv 2 \sqrt3/(\tau_0 \sqrt{g})$.  
$\phi_{+}(t)$ oscillates between $a K$ and $a k K$ with period 
$T_k \equiv 2\tau_0 K'/K$, where $K'\equiv K(k')$. 
For $++$ boundary conditions, but with $r>r_c$: 
\beq
 \mbox{[H]} \quad
\left\{
\begin{array}{l}
\phi_{+}(t) = a\; k'\,K\, \cn(K t/\tau_0; k')\,,\\
r/|r_c| = (2K/\pi)^2 (2 k^2 - 1) > -1\,.
\end{array}
\right. 
 \label{ppHT}
\eeq
$\phi_{+}(t)$ oscillates with period
$2 T_k$ between $- a k' K$ and $a k' K$.  
The qualitative change in the evolution below and above the boundary point occurs at $r=r_c$ via an
exponential relaxation
\beq
\phi_{+}(t)\propto [{\rm cosh}(\pi t/2\tau_0)]^{-1}
\eeq
which is similar to the result at criticality in one dimension\footnote{We mention that from the mathematical standpoint, anharmonic oscillations in one spatial dimension are absent 
because for critical phenomena in a two-dimensional film geometry the Jacobi functions which characterize the mean-field expressions encountered so far for the order parameter profile are replaced by trigonometric functions (see, e.g., \cite{bu-98}).
The analytic continuation of the latter involve hyperbolic functions, which result in an exponential relaxation towards the corresponding asymptotic values. Physically, this is a consequence of the absence of ordered phases in classical one-dimensional systems and therefore the two low-temperature regions disappear.}
and is characterized by a decay time $\tau_{\rm exp} \equiv 2 \tau_0/\pi$. Upon approaching $r_c$, 
\beq
T_k(r\!\to\! r_c) \simeq \tau_{\rm exp} \ln(24/|1-r/r_c|)
\label{eq:divT}
\eeq 
and therefore the periods $T_k$ and $2 T_k$ of the oscillations \reff{ppLT} and \reff{ppHT} --- respectively above and below $r_c$ --- diverge logarithmically with a prefactor set by the exponential decay time $\tau_{\rm exp}$.
For DD boundary conditions we have instead
\beq
\mbox{[H]} \quad
\phi_{D}(t) = 0\, \qquad \mbox{for}\qquad
r  > r_c\,,
\label{DDHT}
\eeq
and
\beq
\mbox{[L]} \quad
\left\{
\begin{array}{l}
\phi_{D}(t) =  a\;K\, \dn(K t/\tau_0 - K';k')\,, \\%[5mm]
r/|r_c| =  - (2K/\pi)^2 (k^2 + 1) < -1\,,
\end{array}
\right. 
\eeq
so that $\phi_{D}(t)$ oscillates with period $T_k$  between $a K$ and $a k K$. 
For $r<r_c$, $\phi_D(t)$ differs from $\phi_+(t)$ only by a half-period shift. 
Figure~\ref{fig:ppTA} shows the period and the amplitude of the oscillations of $\phi_+$
whereas in Fig.~\ref{fig:prof} we present the three typical non-vanishing evolutions.
%
%
%%%%%%%%%%%%%%%%%%%%%%%%%%%%%%%%%%%%%
%%%%%%%%%%%%%%%%%%%%%%%%%%%%%%%%%%%%%
%%%%%%%%%%%%%%%%%%%%%%%%%%%
%%%%%%%%%%%%%%%%%%%%%%%%%%%
%
%
%
%
In the case of  $++$ boundary conditions a mean-field analysis of the critical evolution ($r=0$) was presented in Ref.~\cite{cc-06}, where anharmonic oscillations were reported. Here we have extended the analysis to the entire phase diagram with generic $r$ and boundary conditions, with a careful characterization of the period of the emerging anharmonic oscillations as a function of the quench parameters. This careful study allowed us to highlight
the appearance of the dynamical mean-field transition and reveal its very nature.
%
%
%%%%%%%%%%%%%%%%%%%%%%%%%%%%%%%%%%%%%%%%%%%%%%%%%%
\section{Discussion} 
Our results depend explicitly on the parameter $\tau_0$, which determines the value of  
$r_c\propto \tau_0^{-2}$. 
In Ref.~\cite{cc-06} it has been argued that $\tau_0$ is proportional
to the spatial correlation length $\xi_0$ in the initial state, the inverse of the 
initial mass gap.
Within the mean-field LG formalism $\xi_0$ is proportional to $|r_0|^{-1/2}$, where $r_0$ is the initial value of $r$ in Eq.~\reff{LGH}. Accordingly, $r\tau_0^2$ is proportional to $r/|r_0|$ %
and therefore $r_c \propto r_0$, but with unknown proportionality factors.
%
%
%
%%%%%%%%%%%%%%%%%%%%%%%%%%%%%%
%%%%%%%%%%%%%%%%%%%%%%%%%%%%%%
\begin{figure}
\centerline{\includegraphics[width=0.48\textwidth]{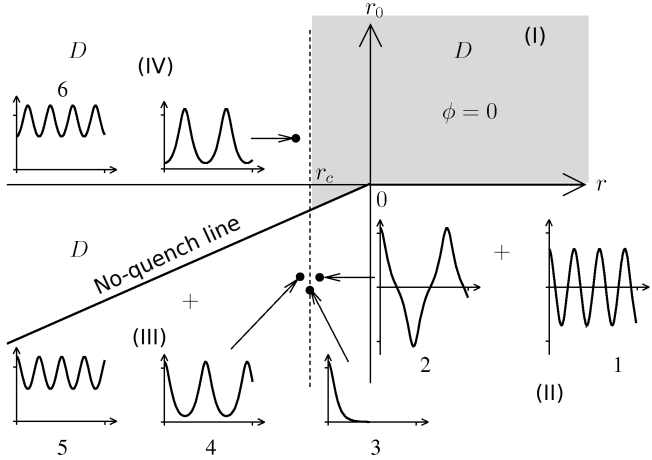}}
\caption{%
Mean-field non-equilibrium phase diagram.
We identify four ``phases'': 
(I) $r>r_c$ and $r_0>0$, where $\phi(t)$ vanishes identically;
(II) $r>r_c$ and $r_0<0$, where $\phi(t)$ shows anharmonic oscillations around zero;
(III) and (IV) where $\phi(t)$ shows anharmonic oscillations with a non-zero average.
These latter two regions are separated by the no-quench line $r=r_0$ and, qualitatively, the corresponding $\phi(t)$ differ only by a half-period shift. 
Upon approaching the 
line $r=r_c$, the period of oscillation diverges according to Eq.~\reff{eq:divT} and $\phi(t)$ decays exponentially to zero with decay time $\tau_{\rm exp}$. % 
}
\label{phasediag}
\end{figure}
%%%%%%%%%%%%%%%%%%%%%%%%%%%%%%
%%%%%%%%%%%%%%%%%%%%%%%%%%%%%%
%
%
Although we cannot fix the exact value of $\tau_0$, the {\it non-equilibrium phase diagram}, which gives the appropriate asymptotic boundary conditions corresponding to a certain choice of the quench parameters  $r_0$ and $r$, can be understood on physical grounds.  
%%%%% DD
In the case of disordered initial state $r_0>0$,
one always expects Dirichlet boundary conditions in the film, independently of the final value $r$.
If such a value exceeds the boundary transition point $r_c$, the order parameter $\phi$ vanishes at all times. 
Conversely, for $r<r_c$, 
one has anharmonic oscillations the features of which depend on $r_0$ through $\tau_0\propto |r_0|^{-1/2}$. 
%%%%% ++
When the quantum system is in an initial state with $r_0<0$, 
the asymptotic boundary conditions depend on the value of $r$. 
From the general features of the profiles in the film (see
Fig.~\ref{fig:profiles}), we expect that if $\phi_0 > \phi_{\rm eq}$,
the corresponding asymptotic boundary conditions is $++$, whereas it is DD for
$\phi_0 < \phi_{\rm eq}$, where  
$\phi_{\rm eq}$ and $\phi_0$ correspond, respectively, to $\phi_b$ and $\phi_s$. 
The  crossover between these two regimes occurs on the {\it no-quench line} $r=r_0$.
All these considerations lead to the non-equilibrium phase diagram  sketched in Fig.~\ref{phasediag} 
(where, in the spirit of our analysis, $r$, $r_0$, and $\tau_0$ are considered as independent parameters).
Remarkably, this phase diagram accounts for all the qualitative 
features found via real-time MF approaches \cite{sb-10,hub}.
In particular, in the Bose-Hubbard model \cite{sb-10}, for  quenches within the ordered (superfluid) phase, 
$\langle\phi^2(t)\rangle$ displays two types of oscillations 
%--- corresponding to "phases" (II) and (III) in Fig.~\ref{phasediag} --- 
separated by the point of "dynamical transition'' at which an exponential relaxation occurs, with decay time $\tau_{\rm exp}$. 
These two different kind of oscillations actually correspond to "phases" (II) and (III) in Fig.~\ref{phasediag}, as it can be realized by comparing the time evolution of the square modulus of the order parameter presented in Fig.~2 (A) and (C) of Ref.~\cite{sb-10} (dashed lines) with the one which can be readily inferred for $\phi^2$ from Fig.~\ref{fig:prof} (a) and (b), respectively. 
Upon approaching the dynamical transition, the period $T_k$ of these oscillation --- expressed in terms of 
the corresponding $\tau_{\rm exp}$ --- was found~\cite{sb-10} to diverge according to the leading term of Eq.~\reff{eq:divT}.
In addition, it turned out that for a fixed initial value $U_0$ of the coupling constant $U$ of the Bose-Hubbard model the dynamical transition occurs at a value $U^d(U_0) = (U_0+U_c)/2$ of the coupling  $U$ after the quench \cite{sb-10},  where  $U_c$ is the value of $U$ at which the model undergoes a Mott-superfluid transition in equilibrium. (The notation here is slightly different from the one used in Ref.~\cite{sb-10}). In terms of the distance $r \propto U-U_c$ from the critical point, the dynamical transition therefore occurs at $r_c \propto U^d(U_0) - U_c = (U_0-U_c)/2 \propto r_0$ where $r_0 \propto U_0-U_c$ is the value of $r$ right before the quench. The fact that $r_c\propto r_0$ provides an additional evidence of the correctness of the picture presented here, which allows us to conclude that 
the point of dynamical transition highlighted in Ref.~\cite{sb-10} is nothing but $r=r_c$, i.e., the shifted critical temperature in the corresponding classical film.
This implies that the dynamical transition is not a MF artifact and therefore it has to be expected generically for $d>1$.
This could also be the case for the dynamical transition points with anharmonic oscillations  
found in the MF evolution of the Hubbard \cite{s-11,hub} and pairing models \cite{bl-06,dyac-07}. 
first, the period of the anharmonic oscillations of the order parameter diverges upon approaching it and, 
second, it separates a region [(I) + (II)] within which the asymptotic time average of the order parameter vanishes from 
a region [(III) + (IV)] within which it does not. 
Fluctuations beyond MF are likely to affect and completely wipe out these oscillations, 
resulting either in a relaxation of the order parameter towards some asymptotic value or even to some
chaotic dynamics in which oscillations (damped or not) with various frequencies are involved, as suggested by the results of Ref.~\cite{sb-10}.
Within the present approach, however, it is difficult to establish whether these fluctuations are relevant only below an upper critical dimensionality (as it is the case of equilibrium critical phenomena) or, instead, they destroy the oscillations in generic dimensionality.
Nonetheless, the RG and scaling analysis of the classical system in the film geometry --- which accounts for critical fluctuations --- suggests that the presence of the dynamical transition is robust in any dimensionality ($\neq1$) and that this transition, as its classical counterpart, is characterized by the second of the two features mentioned above.  
However, the critical-point shift $r_c$ is affected by fluctuations and indeed it scales as $r_c\sim \tau_0^{-1/\nu}$,
where $\nu$ is the equilibrium critical exponent governing the divergence of the order parameter correlation length $\xi$ of the $d+1$-dimensional classical system without boundaries. This fact provides an additional example \cite{dgp-10,gs-11} of how the dynamical behavior after a quantum quench may display some of the universal features of the corresponding classical equilibrium model.
%
%
%
%%%%%%%%%%%%%%%%%%%%%%%%%%%%%%%%

\acknowledgments 
We thank J. Cardy, R. Fazio and A. Silva for  discussions. 
PC thanks the Max-Planck-Institut f\"ur Metallforschung in Stuttgart and SISSA in Trieste for hospitality. 
AG is supported by MIUR within the program "Incentivazione alla mobilit\`a di studiosi
stranieri e italiani residenti all'estero."

\end{document}